# Resonant multiple-phonon absorption causes efficient anti-Stokes photoluminescence in $CsPbBr_3$ nanocrystals

*Zhuoming Zhang[1†], Sushrut Ghonge[2†], Yang Ding[1], Shubin Zhang[2], Mona Berciu[3,4*], Richard D. Schaller[5,6], Boldizsár Jankó[2*], Masaru Kuno[1,2*]*

1. Department of Chemistry and Biochemistry, University of Notre Dame, 251 Nieuwland Science Hall, Notre Dame, IN 46556, United States
2. Department of Physics and Astronomy, University of Notre Dame, 225 Nieuwland Science Hall, Notre Dame, IN 46556, United States
3. Department of Physics and Astronomy, University of British Columbia, Vancouver Campus 325-6224, Agricultural Road, Vancouver BC V6T 1Z1, Canada
4. Stewart Blusson Quantum Matter Institute, University of British Columbia, Vancouver, British Columbia, V6T 1Z4 Canada
5. Department of Chemistry, Northwestern University, Evanston, IL 60208
6. Center for Nanoscale Materials, Argonne National Laboratory, Lemont, IL 60439, United States

† contributed equally

E-mail: berciu@phas.ubc.ca, bjanko@nd.edu and mkuno@nd.edu

ABSTRACT. Lead-halide perovskite nanocrystals such as $CsPbBr_3$, exhibit efficient photoluminescence (PL) up-conversion, also referred to as anti-Stokes photoluminescence (ASPL). This is a phenomenon where irradiating nanocrystals up to 100 meV below gap results in higher energy band edge emission. Most surprising is that ASPL efficiencies approach unity and involve single photon interactions with multiple phonons. This is unexpected given the statistically disfavored nature of multiple-phonon absorption. Here, we report and rationalize near-unity anti-Stokes photoluminescence efficiencies in $CsPbBr_3$ nanocrystals and attribute it to resonant multiple-phonon absorption by polarons. The theory explains paradoxically large efficiencies for intrinsically disfavored, multiple-phonon-assisted ASPL in nanocrystals. Moreover, the developed microscopic mechanism has immediate and important implications for applications of ASPL towards condensed phase optical refrigeration.

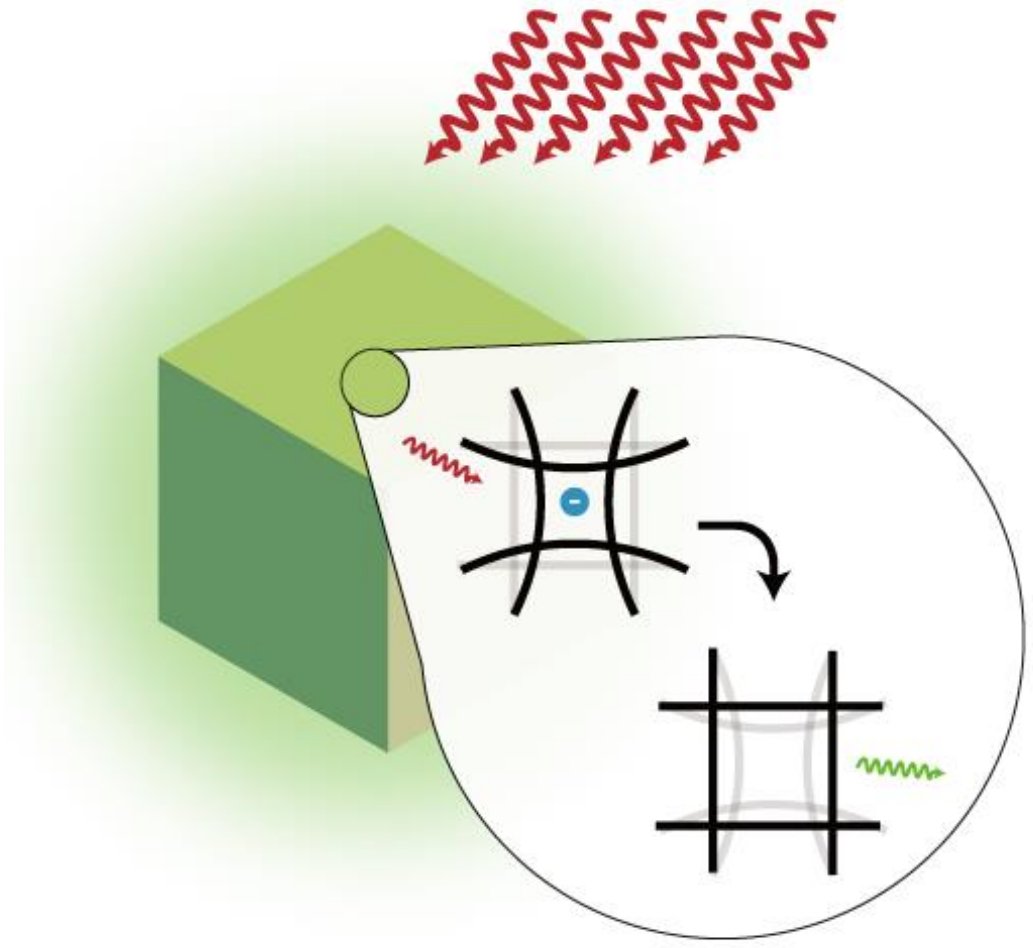

TOC graphic

Unlike most other materials, lead halide perovskites, a new class of materials, interact with light differently in two important ways. First, they readily exhibit near-unity photoluminescence (PL) quantum yields (QYs), where QY is defined as the ratio of emitted photons to absorbed photons. A QY of less than unity indicates non-radiative relaxation processes in the material that lead to heating. Second, they show efficient anti-Stokes photoluminescence (ASPL), where emitted photons have greater energy than those absorbed. The ASPL energy difference, $\Delta E$, represents heat extracted from the material. Making lead halide perovskite nanocrystals special are near-unity ASPL efficiencies, $\eta_{ASPL}$, defined as the fraction of absorbed photons re-emitted at higher energy. A material that simultaneously possesses near-unity QYs and large $\eta_{ASPL}$-values can cool when irradiated with light.

Condensed phase optical refrigeration can, in principle, be achieved via ASPL with important practical applications in vibration-free cryocoolers[1,2,3] and radiation-balanced lasers[4,5]. While optical refrigeration of solids was first conceived by Pringsheim[6] in 1929, the phenomenon has been demonstrated only recently[7] with rare earth-doped glasses/crystals cooled from room temperature to ~91 K.[8] Going to lower temperatures requires optically refrigerating semiconductors.[9,10]

To this end, cesium lead tribromide ($CsPbBr_3$) perovskite nanocrystals (NCs) are intriguing materials for demonstrating semiconductor optical refrigeration.[11] This stems from their soft lattice and strong electron-phonon coupling, which readily leads to polaron formation.[12,13,14,15] More importantly, $CsPbBr_3$ NCs exhibit near-unity, ASPL efficiencies, with previously reported $\eta_{ASPL}$-values of $\eta_{ASPL}$~75% ($\eta_{ASPL}$~32%) for a $\Delta E$=24 meV (102 meV) detuning into the gap.[16] Absorption of multiple phonons by photoexcited carriers is hypothesized to be responsible for large ASPL efficiencies.[11,17] However, given longitudinal optical (LO) phonon energies, ranging from 4-44 meV [See Supporting Information (**SI**), **Table S1**],[18,19,20,21,22,23,24] 1-25 LO phonons must be involved in the up-conversion process.

Multiple-phonon processes are responsible for intra- and inter-band relaxation in semiconductors.[25,26,27] They dictate whether phenomena such as hot carrier extraction[28,29] and phonon bottlenecks[30,31] are observed. Whereas multiple-phonon emission is efficient, the opposite is not. Multiple-phonon absorption requires fast multiple-phonon interactions with photogenerated carriers. Absorption rates increase with electron-phonon coupling energies and decrease exponentially with the number of phonons involved. This makes multiple-phonon absorption processes exceedingly rare.

Despite this, multiple-phonon absorption can be experimentally observed via single-photon/multiple-phonon anti-Stokes photoluminescence (ASPL).[16,32,33] The process entails photoexciting semiconductors below their optical gap whereupon interacting with lattice phonons raises the energy of incident excitations to the band edge. Subsequent band gap photoluminescence confirms multiple-phonon-assisted up-conversion. Despite active efforts in the area[17,34,35,36,37,38,39], an open question remains whether near-unity ASPL efficiencies are possible given intrinsically disfavored multiple-phonon absorption rates.

**Results/Discussion**

We now posit the mechanism for efficient ASPL and the origin of near-unity $\eta_{ASPL}$-values in $CsPbBr_3$ NCs. Visible photothermal absorption (PA) as well as Stokes-photoluminescence (PL) and ASPL spectroscopies are used to map $\eta_{ASPL}$ across a wide range of temperatures and below gap detuning energies. Observed $\eta_{ASPL}$-values are near-unity as well as non-Arrhenius. They are rationalized using a resonant multiple-phonon absorption model, developed below.

Quaternary ammonium passivated $CsPbBr_3$ NCs were synthesized via literature procedures.[40,41] Benzoyl bromide was introduced to pre-made Cs-Pb-oleate to trigger NC growth. Particles were subsequently passivated with didodecyl dimethyl ammonium bromide (DDDMABr), washed with ethyl acetate, and redispersed in toluene. A $CsPbBr_3$ NC/polystyrene matrix (75000 MW, 5% w/w in toluene) was prepared and drop cast onto sapphire for optical measurements. Details of the sample preparation can be found in the Materials and Methods section.

For representative $CsPbBr_3$ NCs (**Figure 1**), an average edge length is $l$=6.8±0.7 nm ($N$=957, **Figure 1a, inset**, and **Figure S1**). A corresponding, room temperature linear absorption spectrum reveals a first excitonic peak at 500 nm and band edge PL at 506 nm ($\lambda_{exc}$ =450 nm, solid green line, 30 meV Stokes-shift[4243]). Accompanying ASPL (open red circles), obtained by exciting below the peak PL energy ($\lambda_{exc}$=535 nm, $\Delta E$=132 meV), reveals little to no difference in observed spectral characteristics. This confirms up-conversion to the same emitting state.

Excitation intensity ($I_{exc}$)-dependent PL measurements show linear behavior over two orders of magnitude in $I_{exc}$. This establishes that observed ASPL involves single-photon/multiple-phonon up-conversion (**Figure 1b**) and agrees with past observations by us[16] and others[44]. Additional measurements show ASPL for energetic detuning up to $\Delta E$=153 meV (**inset**, **Figure 1b**). Beyond this, ASPL arises from two-photon up-conversion, as observed via quadratic $I_{ASPL}$ versus $I_{exc}$ dependencies. See **Figure S2** for more information.

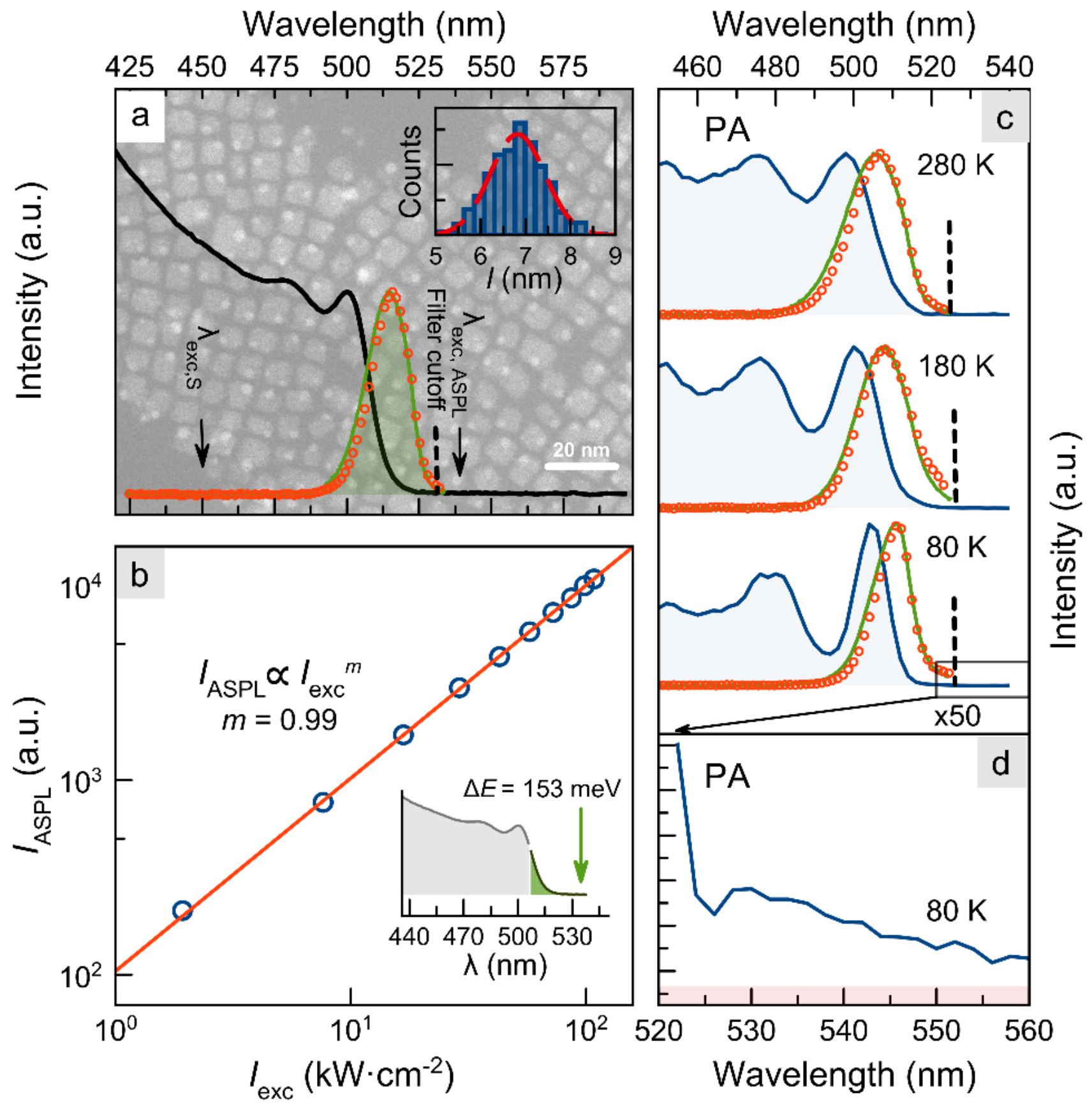

**Figure 1**. **Experimental $CsPbBr_3$ NC ASPL data**. (a) $CsPbBr_3$ NC ($l$=6.8±0.7 nm) ensemble linear absorption (black), Stokes PL (green, $\lambda_{exc}$=450 nm), and ASPL (open red circles, $\lambda_{exc}$=535 nm) spectra. Representative STEM image with associated size distribution histogram (inset). (b) $I_{exc}$-dependent $I_{ASPL}$ ($\lambda_{exc}$=535 nm, $\Delta E$ = 153 meV). Solid red line, linear fit. Inset: Excitation range for single-photon/multiple-phonon up-conversion (green fill). (c) PA (solid blue line/fill), PL (solid green line), and ASPL (open red circles) spectra at 280 K, 180 K and 80 K. Vertical dash lines, filter cutoff. (d) Anti-Stokes PA spectrum at 80 K, baseline shaded red.

Detailed $\eta_{ASPL}$ dependencies with temperature ($T$) and $\Delta E$ are investigated to establish mechanistic insight into underlying multiple-phonon interactions. Whereas $\eta_{ASPL}$ has previously been estimated using experimental absorptance and $I_{exc}$-values that match PL and ASPL intensities[16], a different approach is used here to more broadly access a wide range of $T$ and $\Delta E$. PL and ASPL spectral pairs are acquired between $T$=80 and 280 K for different Stokes (anti-Stokes) detuned energies $\Delta E_S$=170-480 meV ($\Delta E$=20-153 meV) above (below) the emitting state. Photothermal absorption (PA)[45,46] measurements are simultaneously conducted to estimate the amount of absorbed above/below gap light. The importance of PA measurements rests with its ability to discriminate against scattering contributions to NC extinction as well as its sensitive nature, which enables low cross section states to be observed. Details about employed PL, ASPL, and PA measurements as well as an experimental schematic can be found in the **Methods** and in **Scheme S1**.

Acquired PL, ASPL, and PA data at three temperatures (280 K, 180 K, and 80 K) are summarized in **Figure 1c**. For clarity PL and ASPL spectra have been normalized to each other. **Figure S3** summarizes all 11 temperatures studied. Acquired PA spectra reproduce band edge excitonic features, first seen in **Figure 1a**. With decreasing temperature, PL, ASPL and PA spectra all line-narrow and redshift slightly. The **SI** summarizes observed PA and PL Varshni relationships, spectral linewidths, and empirical $T$-dependent peak absorption/peak emission energy differences.[47] Analogous PL and ASPL behavior are observed for all $\Delta E_S$ and $\Delta E$ employed (**Table S2** and **Figure S4**).

**Figure 2a** now shows relevant absorption and emission processes used to model experimental PL ($I_S$) and ASPL ($I_{AS}$) intensities as well as above/below gap PA. Explicit model expressions for $I_S$ and $I_{ASPL}$ are

$$I_S \propto \mathrm{QY}\left(I_{exc,S} A_S\right) \tag{1}$$

and

$$I_{\mathrm{AS}} \propto \eta_{\mathrm{ASPL}} \mathrm{QY}\left(I_{\mathrm{exc,AS}} A_{\mathrm{AS}}\right) \quad (2)$$

where QY is the specimen's emission quantum yield and $I_{\mathrm{exc,S}}$ ($I_{\mathrm{exc,AS}}$) is the Stokes (anti-Stokes) excitation intensity with $A_{\mathrm{S}}$ ($A_{\mathrm{AS}}$) associated absorptances. Corresponding Stokes and anti-Stokes PA expressions are

$$\mathrm{PA_S} \propto \left(I_{\mathrm{exc,S}} A_{\mathrm{S}}\right)\left[\Delta E_{\mathrm{S}} + (1 - \mathrm{QY}) \cdot E_{\mathrm{g}}\right] \quad (3)$$

and

$$\mathrm{PA_{AS}} \propto \left(I_{\mathrm{exc,AS}} A_{\mathrm{AS}}\right)\left[(1 - \eta_{\mathrm{ASPL}} \mathrm{QY}) E_{\mathrm{g}} - \Delta E\right]. \quad (4)$$

The latter PA equations are energy conserving and reflect heat generation due to non-radiative relaxation that follows absorption. More details can be found in the **SI**.

**Equations 1-4** enable $\eta_{\mathrm{ASPL}}$ to be extracted as functions of $T$ and $\Delta E$ via a function, $R$, defined in terms of the ratio of $I_{\mathrm{ASPL}}/\mathrm{PA_{AS}}$ to $I_{\mathrm{S}}/\mathrm{PA_S}$, *i.e.*,

$$R(\Delta E_S, \Delta E, \eta_{\mathrm{ASPL}}, \mathrm{QY}) = \frac{\eta_{\mathrm{ASPL}} \cdot \left[E_{\mathrm{g}} \cdot (1 - \mathrm{QY}) + \Delta E_{\mathrm{S}}\right]}{E_{\mathrm{g}} \cdot (1 - \eta_{\mathrm{ASPL}} \cdot \mathrm{QY}) - \Delta E}. \quad (5)$$

$R$ depends linearly with $\Delta E_{\mathrm{S}}$ for fixed $\Delta E$, allowing $\eta_{\mathrm{ASPL}}$ and QY to be extracted via global fits of obtained $R$ for fixed $\Delta E$ at different $T$. Details of the procedure can be found in the **Methods** and in the **SI**.

**Figure 2b** shows experimental $R$ for $\Delta E$=23 meV (red), $\Delta E$=33 meV (blue), and $\Delta E$=46 meV (green). **Figure S5** (**Table S3**) summarizes $R$ ($\eta_{\mathrm{ASPL}}$) for all $\Delta E$. Associated linear fits yield QY=0.84 and $\eta_{\mathrm{ASPL}}$-values that decrease non-linearly from $\eta_{\mathrm{ASPL}}$ = 0.75 ($\Delta E$=23 meV) to $\eta_{\mathrm{ASPL}}$=0.32 ($\Delta E$=102 meV). **Figure 2c** summarizes this behavior. Note that the fit-extracted QY is in good agreement with specimen QY values, obtained using two independent approaches. The first involves power-dependent photoluminescence measurements of NC films to yield QY=0.84 (**Figure S7a**).[48,49] The second uses an integrating sphere on a colloidal NC suspension to yield QY=0.87 (**Figure S7b**). Details of these measurements can be found in the **SI**.

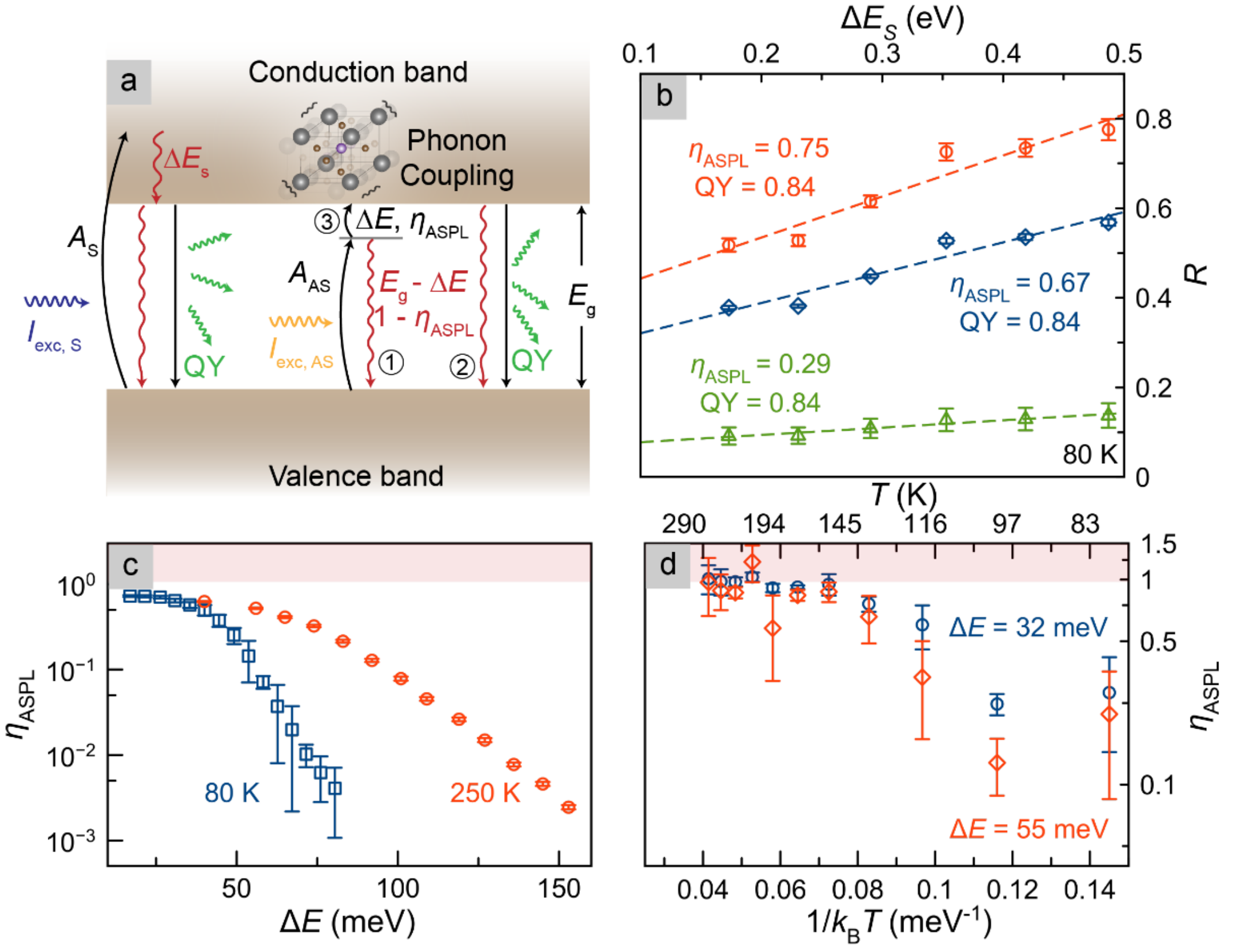

**Figure 2**. **CsPbBr3 NC $\eta_{ASPL}$ temperature and excitation energy de-tuning dependencies.** (a) Schematic of relevant PL, ASPL, and PA processes. (b) $R$ as a function of $\Delta E_S$ for $\Delta E$=23 meV (red circle), $\Delta E$=33 meV (blue diamond), and $\Delta E$=46 meV (green triangle). (c) Extracted $\eta_{ASPL}$ $\Delta E$-dependency at 80 K (open blue squares) and 250 K (open red circles), $\eta_{ASPL}$=1 shadowed red. (d) Extracted $\eta_{ASPL}$ $T$-dependency for $\Delta E$=32 meV (open blue circles) and 55 meV (open red diamonds), $\eta_{ASPL}$=1 shadowed red.

**Figure 2d** shows $\eta_{ASPL}$ $T$-dependencies with data plotted versus $1/k_BT$. **Figure S5** (**Table S4**) summarizes $R$ ($\eta_{ASPL}$) for $T$=80-280 K for $\Delta E$=32 meV and $\Delta E$=55 meV. With increasing $T$, $\eta_{ASPL}$ approaches unity. Log[$\eta_{ASPL}$] does not, however, increase linearly as expected of Arrhenius behavior. This is noteworthy given previous reports suggesting $\eta_{ASPL} \propto e^{-\frac{\Delta E}{k_BT}}$ for both CdS nanobelts[50] and $CsPbBr_3$ NCs.[44] **Figures 2b** and **2c** point to non-Arrhenius $\eta_{ASPL}$.

Up-conversion is now modeled as an interacting two-level system, consisting of valence band-edge $|v\rangle$ and conduction band-edge $|c\rangle$ states, coupled to a phonon mode with energy $\omega_p$. Electron-phonon coupling is described using $g|c\rangle\langle c|(b^\dagger + b)$, where $b^\dagger$ and $b$ are phonon creation and annihilation operators, and $g$ is the electron-phonon coupling constant (**Table S1**). The effective Hamiltonian, $H$, acting on the electron Hilbert space and phonon Fock space, is given by

$$H = E_v|v\rangle\langle v| + E_c|c\rangle\langle c| + \omega_p b^\dagger b + g|c\rangle\langle c|(b^\dagger + b). \tag{6}$$

Eigenstates consist of polarons (conduction band states dressed with $n$ phonons) and similarly dressed valence band states,

$$|c_n\rangle = |c\rangle \otimes \frac{(B^\dagger)^n}{\sqrt{n!}}|\tilde{0}\rangle \quad \text{and} \quad |v_n\rangle = |v\rangle \otimes \frac{(b^\dagger)^n}{\sqrt{n!}}|0\rangle. \tag{7}$$

$B^\dagger$ is the dressed-phonon creation operator, $B^\dagger = b^\dagger + g/\omega_p$, with a coherent vacuum state, $|\tilde{0}\rangle = e^{-\frac{g^2}{2\omega_p^2} - \frac{g}{\omega_p}b^\dagger}|0\rangle$. Corresponding conduction (valence) band energies are $E(c_n) = E_c + n\omega_p - g^2/\omega_p$ $[E(v_n) = E_v + n\omega_p]$ with an associated polaron energy of $E_c - \frac{g^2}{\omega_p}$.

At thermal equilibrium, the probability of occupying a $n$-phonon state is $e^{(-n\omega_p/k_BT)}/Z$, where $Z$ is the partition function. Using the Kramers-Heisenberg equation for resonant phonon absorption[51], the $N$-phonon absorption rate is

$$w_N = \left|\sum_{n=0}^{\infty} \frac{\langle N|\tilde{n}\rangle\langle\tilde{n}|0\rangle}{n\omega_p + i\zeta}\right|^2 \delta(N\omega_p - \Delta E) \tag{8}$$

where $\zeta$ is the conduction band recombination rate and $|\tilde{n}\rangle = [(B^\dagger)^n/\sqrt{n!}]|\tilde{0}\rangle$. In turn, the up-conversion rate, $w_{\text{uc}}$, is the weighted sum of phonon absorption rates with all possible initial states, *i.e.*, $w_{\text{uc}} = \sum_N (e^{-N\omega_p/k_BT}/Z) w_N$. Keeping only the resonant term ($n = 0$) and substituting $N = \frac{\Delta E}{\omega_p}$ from the Dirac delta function, yields

$$w_{\mathrm{uc}} \propto \frac{e^{-\frac{\Delta E}{k_B T}}}{\sqrt{\Delta E}} \exp\left[\frac{\Delta E}{\omega_p}\left(1 - \log\frac{\Delta E \omega_p}{g^2}\right)\right], \tag{9}$$

which is non-Arrhenius, thus rationalizing observed non-Arrhenius behavior in **Figures 2c** and **2d**.

**Figure 3** compares experimental $\eta_{\mathrm{ASPL}}$ with **Equation 9** where $\eta_{\mathrm{ASPL}}(T, \Delta E) = \frac{w_{\mathrm{uc}}(T,\Delta E)}{w_{\mathrm{r}}}$ with $w_{\mathrm{r}} = w_{\mathrm{uc}} + w_{\mathrm{nr}}$ the total depopulation rate, including nonradiative recombination. Evident is that resonant multiple-phonon absorption by polarons quantitatively accounts for $\eta_{\mathrm{ASPL}}$ $\Delta E$- and $T$-dependencies. Extracted electron-phonon coupling and phonon energies are: 80 K, $g$=21.4±0.34 meV, $\omega_p = 5.17 \pm 0.27$ meV; 250 K, $g$=41.1±0.58 meV, $\omega_p = 23.4 \pm 2.2$ meV. Model-extracted $\omega_p$ and $g$ are in good agreement with LO phonon and electron-LO phonon coupling energies previously reported for $CsPbBr_3$ (**Table S1**).

Ab-initio calculations[23,24] show phonons with energies 4-7 meV (Cs vibrations in the Cs-Br plane and rotation of $PbBr_6$ octahedra), 9.5 meV (Cs-Br in-plane motion), and 18-19 meV (Pb-Br stretching mode in the $PbBr_2$ plane). Observed $T$-dependent differences in $\omega_p$ and possibly $g$ may reflect variations to the dominant phonon mode, given changes to the Bose-Einstein distribution at 80 K ($k_{\mathrm{B}}T \approx 7$ meV) and 250 K ($k_{\mathrm{B}}T \approx 22$ meV). Direct identification of involved phonon modes would require observing an isotope effect wherein $\omega_p$ scales with the inverse square root of the isotope mass.

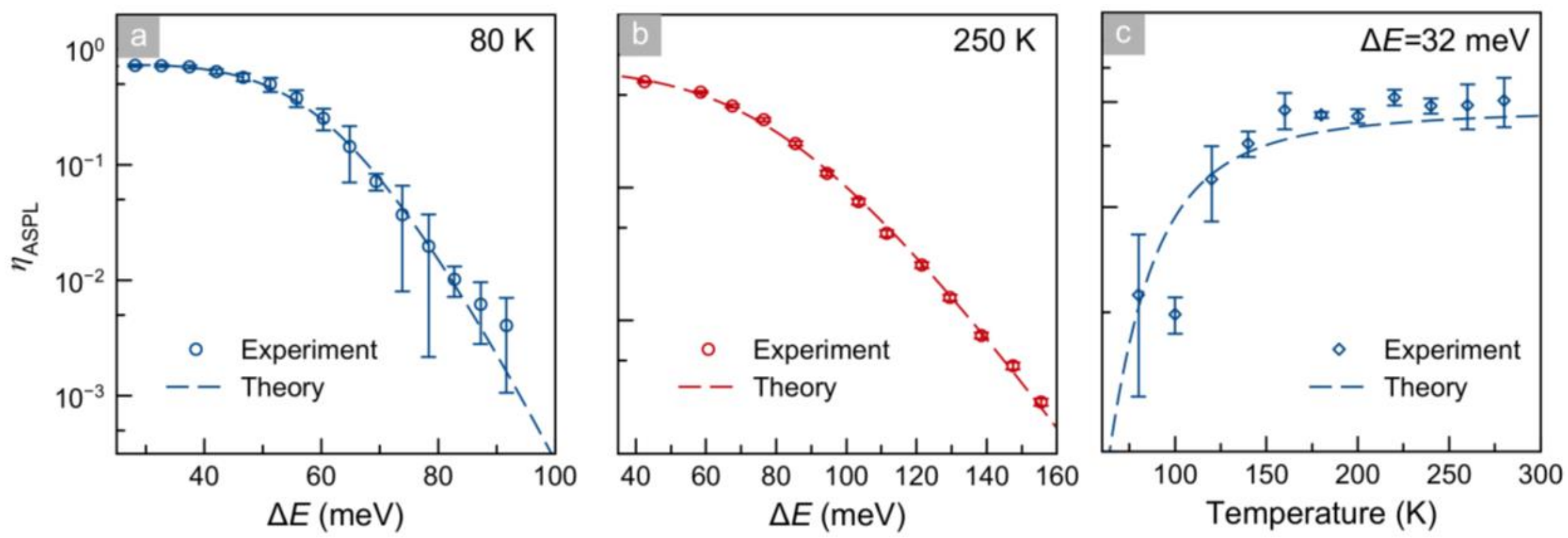

**Figure 3**. **CsPbBr3 NC $\eta_{\text{ASPL}}$ comparison between experiment and theory**. Model (dashed lines) and experimental (symbols) $\eta_{\text{ASPL}}$ $\Delta E$ dependency comparison at (a) 80 K and (b) 250 K. (c) Analogous $\eta_{\text{ASPL}}$ $T$-dependent comparison for fixed $\Delta E$=32 meV.

Although the model does not explicitly predict numerical values for up-conversion rates, given that $\eta_{\text{ASPL}}$ depends on both $w_{\text{uc}}$ and $w_{\text{nr}}$, we performed ultrafast transient up-conversion absorption experiments using a 5 kHz amplified Ti:S laser with an 80 fs pulsewidth to estimate $w_{\text{uc}}$. Results from these measurements with NCs excited below gap at varying $\Delta E$ (**Figure S8** and **Table S5**) suggest that up-conversion timescales range from 52-220 fs for $\Delta E$=33-100 meV. Given $\eta_{\text{ASPL}}$ in the range $\eta_{\text{ASPL}}$=0.2-0.8, $1/w_{\text{nr}}$ ranges from 13-880 fs. This points to multiple-phonon absorption becoming competitive with non-radiative relaxation.

Our model implies that the polaron binding energy $\left(\frac{g^2}{\omega_p}\right)$ contributes to the Stokes shift (*i.e.*, the difference in energy between band edge absorption and emission peaks). Originally, a confined hole state model predicted NC Stokes shifts of sizable magnitude.[43] More recent DFT calculations, which directly included spin-orbit coupling effects, however, revealed that the confined hole state transition was not dark.[42] This prompted further efforts to find an explanation for perovskite NC Stokes shifts. While band edge exciton fine structure from both spin-orbit coupling and the electron-hole exchange interaction does split ground state excitons into dark (singlet) and bright (triplet) states[52,53] , observed dark/bright exciton splittings are much too small (<10 meV) to rationalize size-dependent 20-80 meV Stokes shifts seen in $CsPbBr_3$ NC ensembles.

The polaron model, which involves phonon-dressed states, preserves exciton fine structure and together with its polaron binding energy may ultimately rationalize the overall magnitude of experimental Stokes shifts. Note though that a full theoretical description of a polaron-based Stokes shift has yet to be developed.

## Conclusion

In summary, we find that efficient, multiple-phonon-assisted ASPL in $CsPbBr_3$ NCs is a consequence of resonant phonon absorption by polarons caused by strong electron-phonon coupling. Resonant absorption makes possible counterintuitive, near-unity $\eta_{ASPL}$ despite sizable sub-gap detuning energies. Moreover, the developed model broadly rationalizes near-ubiquitous, single-photon/multiple-phonon PL up-conversion in semiconductor NCs, identifies resonant multiple-phonon absorption by polarons as the origin of $CsPbBr_3$ NC up-conversion, points to simultaneous near-unity PL quantum yields and near-unity $\eta_{ASPL}$-values being self-consistent, and possibly leads to rational approaches for discovering and engineering materials capable of efficient anti-Stokes photoluminescence.

## Methods/Experimental

Chemicals. Lead (II) acetate trihydrate ($PbOAc_2 \cdot 3H_2O$, 99.99%), cesium carbonate ($Cs_2CO_3$, 99%), 1-octadecene (ODE, 90%), didodecyl dimethyl ammonium bromide (DDDMABr, 98%), ethyl acetate (EtOAc, 98.8%), toluene (anhydrous, 99.6%), and polystyrene (PS, 75000 MW) were purchased from Sigma Aldrich. Oleic acid (OA, 99%), benzoyl bromide (BzBr, 97%), and didodecylamine (DDAm, 97%) were purchased from TCI. All chemicals were used as received.

$CsPbBr_3$ NC synthesis and purification. $CsPbBr_3$ NCs were synthesized using a modified literature procedure.[40] In brief, a Cs-Pb cationic stock solution was prepared by mixing $Cs_2CO_3$ (760 mg, 2 mmol) and $PbOAc_2 \cdot 3H_2O$ (160 mg, 0.50 mmol) in a three-neck flask, followed by injection of 10 mL of OA. The mixture was then degassed under a pressure of < 20 mbar on a Schlenk line for three hours at 100 °C to produce Cs-Pb-oleate and remove carbonic and acetic acid byproducts. The resulting product was stored under nitrogen. An accompanying BzBr stock solution was prepared by dissolving 1.0 mL (8.5 mmol) of BzBr in 10 mL of toluene. A surface passivating DDDMABr stock solution was prepared by dissolving 0.231 g (0.5 mmol) of DDDMABr in 20 mL of toluene.

$CsPbBr_3$ NCs were synthesized in a nitrogen atmosphere glovebox. 750 μL of the Cs-Pb stock solution was mixed with 5 mL of ODE and 221.5 mg (0.63 mmol) of DDAm in a 20 mL scintillation vial. The vial was inserted into an aluminum block, set atop a magnetic stir plate. The Al block's temperature was then raised to values between 70 °C and 110 °C, depending on desired NC size. Once the final growth temperature was reached, the vial was removed whereupon 0.55 mL of the BzBr stock solution was injected to initiate NC growth. An immediate color change from clear to yellow was observed. The reaction was subsequently allowed to cool to room temperature.

After cooling, 2 mL (50 mmol) of a surface passivating DDDMABr stock solution was added to 3 mL aliquots of as-prepared NCs to replace their original oleate ligands. An immediate photoluminescence brightening of ensembles results. 16 mL of EtOAc was subsequently added followed by centrifugation at 4400 rpm to recover a NC precipitate. Recovered NCs were resuspended in neat toluene for subsequent optical measurements. Samples were filtered through 0.22 μm PTFE filters to remove any particulates.

Characterization. Scanning transmission electron microscopy (STEM) images were acquired with a Spectra 300 microscope (Thermo Scientific), operating an accelerating voltage of 300 kV. Samples were prepared by drop-casting dilute NC solutions in toluene onto ultrathin amorphous carbon substrates with copper supports (Ladd). ImageJ was used to estimate NC edge lengths via its FIJI Particle Analysis Procedure. NC edge lengths were then obtained as square roots of obtained areas. NC shapes are implicitly assumed to be cubic.

Optical sample preparation. A polystyrene (PS) stock solution was prepared by dissolving 50 mg of PS in 950 mg toluene (5% w/w PS solution in toluene). The mixture was stirred at 50 °C until clear. For optical measurements, 1 mL aliquots of NC ensembles were mixed with an equivalent volume of the PS stock solution. The resulting mixture was then drop-cast onto 1 mm thick sapphire substrates (Esco Optics).

Visible photothermal absorption spectroscopy. A fiber-based supercontinuum laser (NKT, SuperK extreme), coupled to an acousto optical filter was used as a tunable, visible wavelength excitation source. The supercontinuum intensity was modulated at 4 kHz using a mechanical chopper (Thorlabs, MC2000). Intensity modulated light was then focused onto NC specimens using a 0.4 numerical aperture (NA) long working distance objective (Mitutoyo).

The specimen was mounted to the cold finger of an optical microscope cryostat (Cryo Industries of America). During measurements, specimen temperatures were varied from 80 K to 280 K with a temperature controller (Lakeshore). The cryostat was actively pumped using a turbo pump with its base pressure maintained at $10^{-5}$ torr.

A continuous-wave, red probe laser (785 nm, Coherent OBIS) was simultaneously focused onto the specimen from the opposite (counterpropagating) direction. A 0.45 NA objective (Nikon) was used to focus the probe light. Pump (average) and probe intensities at the sample were 22 kW $cm^{-2}$ and 15 MW $cm^{-2}$ respectively.

Scattered probe light was collected using the same Nikon objective and was directed onto a photodiode (Thorlabs, PDA36A). The photodiode signal was then fed into a lock-in amplifier (Stanford Research Systems, SR 830), referenced to the mechanical chopper. A typical lock-in integration time was 30 ms.

PA spectra were acquired by scanning the excitation laser wavelength while recording the lock-in signal. Ten spectra were acquired and averaged. Data acquisition and processing were carried out using home-written Python software. More information about the photothermal technique can be found in Reference 45.

Ensemble PL and ASPL spectroscopy. Ensemble PL and ASPL spectra were acquired using the same microscope employed for photothermal absorption measurements. The output of the supercontinuum was focused onto specimens and induced PL or ASPL was collected with a 0.45 NA (Nikon) long working distance objective. Both PL and ASPL were filtered using dielectric filters (Semrock) to remove any excitation light. Spectra were acquired using a spectrometer (Acton SpectraPro, SP-2300, 150 groove/mm, 800 nm blaze), coupled to an electron multiplying CCD camera (Andor, Ixon Ultra). Detuned PL (ASPL) spectra were acquired by changing the excitation wavelength $\Delta E_S$ ($\Delta E$) above (below) the energy of the emitting state. Ten PL or ASPL spectra were acquired, integrated between 400-600 nm, and averaged. All data were acquired and processed using purpose-built Python software.

## Acknowledgements

This work is dedicated to the memory of Mansoor Sheik-Bahae. We thank the MURI:MARBLe project under the auspices of the Air Force Office of Scientific Research (Award No. FA9550-16-1-0362) for financial support. This work was also supported, in part, by the National Science Foundation under award DMR-1952841. MB acknowledges support from the Max Planck-UBC-UTokyo Center for Quantum Materials and Canada First Research Excellence Fund (CFREF) Quantum Materials and Future Technologies Program of the Stewart Blusson Quantum Matter Institute and the Natural Sciences and Engineering Research Council of Canada.

## Author Contributions

Z. Z. and S. G. contributed equally to this work.

## Notes

The authors declare no competing financial interest.

ASSOCIATED CONTENT

**Supporting Information**.

The Supporting Information is available free of charge at https://pubs.acs.org/

Literature-reported $CsPbBr_3$ phonon energies and associated electron-phonon coupling energies, representative STEM images of obtained $CsPbBr_3$ NCs, $CsPbBr_3$ NC ensemble linear absorption, PL spectra, and power-dependent ASPL at different wavelengths, schematic of instrumentation used to conduct PL, ASPL, and PA spectroscopies, PA and PL spectra of $CsPbBr_3$ ensemble at different temperatures, table summarizing *T*-dependent absorption and peak PL energies along with corresponding full width at half maxima (FWHM) and Stokes shifts, $CsPbBr_3$ NC *T*-dependent PL FWHM and Stokes shifts, mathematical model for estimating $\eta_{ASPL}$ based on PA, PL, and ASPL intensities, details of extractions of $\eta_{ASPL}$ at different temperatures and different detuning energies, summary of $\Delta E$-dependent $\eta_{ASPL}$ at $T$=80 K and $T$=250 K, summary of *T*-dependent $\eta_{ASPL}$ for $\Delta E_S$ =32 meV and $\Delta E_S$ =55 meV, details of QY measurements, details of femtosecond transient absorption up-conversion spectra of $CsPbBr_3$ NCs, PL and ASPL spectra from a $CsPbBr_3$ NC suspension.

[53] Sercel, P. C.; Lyons J. L.; Wickramaratne, D.; Vaxenburg, R.; Bernstein, N.; Efros, A. L. *Nano Letters* **2019** *19* (6), 4068-4077.